# A protocell design for bioaccumulation applications

**Ian von Hegner**

**ABSTRACT** This article provides a review of specific example of recombinant cell and protocell technology, moving from what is presently known to suggesting how novel application of existing methodologies could be utilized to design a complex synthetic system in form of a self-sufficient light empowered protocell. A practical application of protocells using a primary example of desalination in water treatment is given, followed by a more general review regarding bioaccumulation and bio-diagnostics, outlining the possibilities associated with applications of protocells. The key hypothesis is that the inside-negative electrochemical membrane potential generated by $Cl^-$ pump activity via halorhodopsin could also be utilized to drive the accumulation of cations into a protocell. Thus, the functional expression of halorhodopsin could energize proton-coupled uptake of substances or metals through a selective cotransport channel for a number of applications in biotechnology, molecular medicine, and water biotechnology.

## INTRODUCTION

Between 4.0-3.5 billion years ago the first cell on Earth emerged. This first cell existed only a brief moment, and represented the beginning of life as we know it [Altermann et al., 2003]. Shortly after the abiogenesis this original cell split in two, and these two split again, and during a short geological time scale Earth was populated by unicellular organisms. That was the beginning of the history of life on this planet, and thus the beginning of the history of biology itself.

Synthetic biology reflects the view that the best way to investigate the accuracy and limits of current biological knowledge and phenomena is to modify or engineer a different artificial version of a complex biological system and compare its functions with theoretical expectations [Solé et al., 2007]. Efforts to understand the origin of life have resulted in the research in protocells. Research in protocells is a bottom-up form of synthetic biology that is synthesizing novel kinds of living systems from parts that have never been alive [Rasmussen et al. 2009].

Creating the first bottom-up protocell will mark the first time science have synthesized life from wholly nonliving materials. Advances in protocell research are gaining public awareness, and this awareness will increase as work on protocells proceeds from the research to the development stage [Mark et al., 2009]. There is much ground to be explored before autonomous protocells become a reality. As well as basic construction issues, design will focus on maintaining compatibility—whether by separation of processes or tailoring of reagents. But when we do gain the ability to create protocells this could have profound impacts on many people's view of life. Funding agencies significantly shape the direction and conduct of research activities, and public acceptance play an important role.

While having vast importance in their own right, protocells also offers great promise in biotechnology, where they offer a valuable simplification compared to the use of for instance recombinant bacteria, which might ease public acceptance [Amezaga et al., 2014]. Carefully selected components of metabolic pathways can hypothetically be utilized to create purpose-built protocells for a variety of applications with highly specific functions. The ability of bacteria to degrade a variety of organic compounds has been used in waste processing and bioremediation and the production of chemicals to mention just a few examples [Ishige et al., 2005]. Their wide applicability could also be utilized by harvesting and expressing membrane transport proteins from i.e. cyanobacteria into the synthetic membranes of protocells. A number of membrane proteins have been successfully integrated into GUVs of a size appropriate for a protocell, including potassium channels [Aimon et al., 2011], and bacteriorhodopsin [Girard et al., 2004]. Energization of desalination by solar light through photosynthetic biological or synthetic machines offers a potential opportunity to utilize biological processes in protocell design [Amezaga et al., 2014].

In this review article a practical application of protocells using a primary example of desalination in water treatment, followed by a more general approach to bioaccumulation and bio-diagnostics will be discussed, outlining the possibilities associated with applications of self-sufficient protocells, many of which have relevance to other biotechnologies.

**RECOMBINANT CELLS AS DESALINATION RESERVOURS**

Progress in recombinant biology techniques in the 1990s heralded new approaches enabling exploration of biological systems in unprecedented detail. The ability to clone, modify and express a huge array of proteins was extended yet further in synthetic biology by the capacity to reconstitute recombinant macromolecules into cellular systems [Barla et al., 2013].

The first and most basic example in this review is a desalination process consisting of the creation of a low-salt biological reservoir system within a water environment that can function as an ion exchanger [Amezaga et al., 2014]. Cyanobacteria are a model photosynthetic prokaryotic organism capable of growing phototrophically due to oxygenic photosynthesis and heterotrophically at the expense of reduced carbon sources in the dark [Nogales et al., 2012]. Because of the process of complementary chromatic adaptation [Bennett et al., 1973], cyanobacteria can utilize a wide spectrum of photosynthetically active radiation as their primary source of energy.

Cyanobacteria can be cultivated under outdoor conditions and a possible usage of cyanobacteria for desalination was the reclamation of saline soils in India and the former Soviet Union using endemic strains [Apte et al., 1997; Singh et al., 2010]. Thus, removal of cyanobacterial mats formed after rainfall also removed salt from the soil. As a line of defense, cyanobacteria employ a range of $Na^+$ export proteins in their cell membrane, where the activation of $Na^+$-export mechanisms such as $Na^+/H^+$-antiporters and $Na^+$/ATPases results in the reduction of the sodium concentration [Nitschmann et al., 1992]. The hydrolysis of ATP by ATPase is the driving force for $Na^+$ transport and the rate of hydrolysis is controlled by ion potential, i.e. the lower the potential, the higher the activity [Wiangnon et al., 2007].

Naturally, if the utilization of ATP is blocked, either by itself for some reason or is deliberately manipulated through genetic or environmental means, the export of sodium would come to a halt. Once active $Na^+$ export has come to a standstill, there will be net $Na^+$ influx into a cell until equilibrium with the external medium is reached [Amezaga et al., 2014]. Further extraction of $Na^+$ from the medium will then require an energy source. If desalination of water is the objective, then it is necessary to prevent renewal of $Na^+$ export, and the energy-harvesting system employed during this phase should not use ATP as an intermediate. Good candidates for ATP-independent light-powered biological batteries are as suggested by Amezaga et al. (2014) halorhodopsin (Hr) proteins, that is, inward-directed $Cl^-$ pumps [Lanyi, 1990]. The Hr from Natronomonas pharaonis (pHr) has been cloned and functionally expressed in heterologous systems such as E. coli [Hohenfeld et al.., 1999].

As will be explained in more detail in the next section, the idea is to introduce such a light empowered rhodopsin pump into the membrane. The electrochemical potential created across the recombinant membrane due to anion transport could be used to energize proton-coupled uptake of $Na^+$ through a selective cotransport protein [Amezaga et al., 2014].

This approach could hypothetically create a recombinant cyanobacteria strain, using novel methods [Heidom et al., 2011; Minas et al., 2015], where the resulting light empowered salt accumulator bypasses the endogenous energy metabolism (photosynthesis and respiration) and should therefore remain functional even when increasing intracellular $Na^+$ levels inhibit other metabolic functions of the host. Expression of Hr in a high-density cyanobacterial culture should remove both $Cl^-$ and $Na^+$ from surrounding seawater and thus provide a means for bioaccumulation [Amezaga et al., 2014].

However, excess accumulation of NaCl in natural and presumably recombinant cells threatens to lead to destabilization of membranes and proteins [Huflejt et al., 1990] and a complete halt of photoautotrophic growth [Bhargava et al., 2003]. Further, modifying existing organisms this way still holds some inherent limitations, the degree of control by re-directing living cells still holds many unknowns and attempts to integrate new biological functions are often unsuccessful. While optimization can be accomplished to some degree through metabolic engineering of existing organisms, e.g., by evolution and selection [Smith et al., 2011] host organism modification is not trivial [Fischer et al., 2008], and a number of issues associated with modifying existing organisms exists.

A top down approach could be attempted to simplify the recombinant cyanobacteria in order to make it a better design. However, theoretically, a protocell designed bottom up offer here a much simpler solution. The degree of control over protocell design far exceeds that which can be obtained by re-directing living cells, which will be discussed shortly.

**THE MEMBRANE POTENTIAL**

As previously explained, once active $Na^+$ export has come to a standstill, there will be net $Na^+$ influx into a cell until equilibrium with the external medium is reached. Further extraction of $Na^+$ from the medium will then demand an energy source. An energy source is vital in any cell, synthetic or living. The latter utilize biochemical respiratory pathways and the electron transfer chain to produce energy currency such as ATP to drive reactions.

In the simplest protocell design, all of the required small molecule components including nucleotides and amino acids can be provided for biosynthesis, with ATP provided as an energy source [Noireaux et al., 2004]. These simple designs eliminate the demand for energy generation and while impressive in their own right, are unlikely to be desirable designs due to the need for continual renewal of all molecules. To prevent renewal of $Na^+$ export, the energy-harvesting system employed during this phase should not use ATP as an intermediate.

An alternative and more sustainable design would require suitable pathways for independent and efficient energy generation essentially designing an autonomous protocell capable of utilizing an easily available energy source such as sunlight to support function. Such a synthetic design would require a straightforward metabolic pathway to harvest the light energy into a usable form of energy [Minas et al., (2015].

Proposed candidates for ATP-independent light-powered biological batteries could be halorhodopsin (Hr) proteins [Amezaga et al., 2014]. Hr is an inward-directed $Cl^-$ pump [Mukohata et al., 1981; Lanyi, 1990]. Hr, functioning as a $Cl^-$ pump, has the ability to transport $Cl^-$ against an electrochemical gradient, and can generate an inside-negative membrane potential to support ATP synthesis [Schobert et al., 1982; Rudiger et al. 1997]. Hrs naturally occur in extremely salt-tolerant archaea (haloarchaea) and are membrane-integral proteins of the rhodopsin superfamily that form a covalent bond with the carotenoid-derived chromophore all-trans-retinal [Klare et al., 2008]. Absorption of a photon with a defined optimal wavelength induces trans-cis isomerization of retinal, which triggers a catalytic photocycle of conformational changes in the protein, resulting in the net import of one chloride per photon into the cytoplasm [Kolbe et al., 2000; Chizhov et al., 2001; Kouyama et al., 2010].

An effective implementation strategy would be to segregate the energy-generating pathway in an artificial protoorganelle, basically mimicking the mitochondria found in eukaryotic cells. In mitochondria the formation of a proton gradient is an important step in energy harvesting and conversion. Although such procedures appear complicated, there is precedent in the form of synthetic systems for ATP production. For example artificial chloroplasts utilizing ATP synthase and bacteriorhodopsin can be designed from a limited number of components [Choi et al., 2005]. Through a sophisticated synthetic biology approach, artificial proton pumps were engineered from modified cytochrome c and cytochrome c oxidase in artificial polymerosomes [Hvasanov et al., 2013]. These proton pumps generated an artificial proton gradient across the polymerosome membrane, which could hypothetically be harnessed for energy production. This could hypothetically be linked to the formation of ATP required in protocells to form the basis of a straightforward light driven energy source.

The halorhodopsin from *Natronomonas pharaonis* (pHr) has been cloned and functionally expressed in systems such as Xenopus laevis oocytes [Gradinaru et al., 2008]. This expression resulted in a light-dependent $Cl^-$-inward current and thus a considerable negative shift in the membrane potential (∼−400 mV), indicating that pHr is a highly active $Cl^-$ pump [Seki et al., 2007]. The opportunity to artificially manipulate a cell's membrane potential through pHr in conjunction with light-activated cation channels (e.g. channel rhodopsin) has been exploited in the field of optogenetics, achieving control of action potentials in nerve cells with potential medicinal applications [Fenno et al., 2011; Zhang et al., 2011].

With the results from Hvasanov et al. 2013, Mukohata et al. 1981, and Zhang et al. 2011 in mind, a hypothesis can be put forward, that the inside-negative electrochemical membrane potential ($V_m$) generated by $Cl^-$ pump activity via Hr could also underpin specialized accumulation of cations, positively charged substances, in protocells [Amezaga et al., 2014]. Thus, the functional expression of Hr could energize proton-coupled uptake of a range of nutrients such as $Ca^{2+}$, $Mg^{2+}$, $K^{+}$, $Fe^{2+}$, or metals such as $Na^{+}$, $Cd^{2+}$, $Ni^{2+}$ for bioaccumulation.

Expression of the electrogenic anion transport via pHr in a protocell should remove both $Cl^{-}$ and $Na^{+}$ from surrounding seawater and thus provide better insight into means for desalination. The observed $K_m$ values of Hrs for chloride uptake (approximately 25 mm for chloride; [Duschl et al., 1990] are in an optimal range for this purpose.

To increase the speed of $Na^{+}$ accumulation, the $Na^{+}$ conductance of the membrane might need to be enhanced by coexpression of $Na^{+}$-selective and voltage-dependent channels or carriers with Hr. Such proteins could be NaChBac, a transmembrane-spanning protein cloned from *Bacillus halodurans* the first functionally characterized bacterial voltage-gated $Na^{+}$-selective channel [Ren et al., 2001] or the voltage-dependent $Na^{+}$ channel $Na_V$PZ from *Paracoccus zeaxanthinifaciens* [Koishi et al., 2004].

**A SYNTHETIC LIGHT DRIVEN PROTOCELL**

The previous discussion could be applied to recombinant bacteria, where a top down approach could be attempted to simplify the recombinant cyanobacteria in order to make it a better design. At present, designer organisms are based on living cells that have undergone modifications to selected metabolic pathways or have had new pathways incorporated [Barla et al. 2013].

The potential benefits in modifying existing organisms this way is attractive. “Housekeeping” metabolic processes, including the ability to generate energy currency, are already present within host organisms and their endogenous biosynthetic pathways can be harnessed to produce precursors for synthetic chemistry. Despite these obvious benefits there are some inherent limitations [Minas et al., (2015]. Even in a seemly simple prokaryotic cell the regulatory network controlling substrate-product relationships is so complex that attempts to integrate new biological functions are often unsuccessful. For instance, modifying an organism’s metabolism decreases its overall fitness, as a result of the buildup of inhibitory precursors [Pitera et al., 2007; Zhu et al., 2002] or from the inherent toxicity of the preferred end products to the host cell. For example yeast cells designed to produce artemisinic acid, a chemical precursor to artemisinin, are characterized by a noticeable increase in indicators of cellular stress responses [Ro et al., 2008].

As an alternative to the top-down approach, which aims to reduce natural biological systems, protocells could offer a superior alternative with their simplification of chassis. Designed from the bottom upwards, protocells comprise elementary systems containing only a minimal complement of components required to execute necessary cellular function(s). So although the processes to maintain life obviously are essential for biological cells, many of these systems can be omitted or considerably simplified in protocell design, and pathways relating to function can be designed to optimize performance. Even with this reduced level of complexity, protocells might still perform a specified function as effectively as biological cells.

The first step will be to purify and express Hr in an appropriate host organism, such as an *E. coli* strain, to produce recombinant functional halorhodopsins along with highly selective ion transporters such as $Na_V$PZ [Amezaga et al., 2014]. Integral membrane proteins are still considered as complicated with regards to their biochemistry and their folding and function is dependent on an optimal membrane environment. Thus, there have until recently only been few examples of well characterized membrane proteins and methodologies for generating functional recombinant membrane proteins.

Progress during the last decade studying membrane protein function in a native membrane environment under controlled conditions has seen the emergence of effective techniques for integration of recombinant proteins into lipid membranes in vitro [Seddon et al., 2004]. It seems reasonable that these same methods could be adapted to the integration of surface transporters and receptors into protocell membranes. A number of membrane proteins have been successfully integrated into GUVs of a size appropriate for a protocell, including potassium channels [Aimon et al., 2011], mechanosensitive channels [Folgering et al., 2004], a $Ca^{2+}$ ATPase and bacteriorhodopsin [Girard et al., 2004]. To date, several Hr proteins from different species have been characterized [Klare et al., 2008; Fu et al., 2012]. The Hr from Natronomonas pharaonis (pHr) has been cloned and functionally expressed in heterologous systems such as E. coli [Hohenfeld et al., 1999].

The next step will be to integrate lipid patches containing the recombinant proteins, extracted from E.coli, into a synthetic cell membrane made of polymersomes. The membrane of polymersomes can be considered as a reservoir system for both hydrophobic and amphiphilic molecules similar to cell membranes, which incorporate cholesterol and membrane proteins. The folding, and therefore function, of membrane proteins is known to be heavily influenced by the biophysical properties of the membrane in which they are embedded [Laganowsky et al., 2014 ; Curran et al., 1999], which are in turn determined by the lipid composition of the membrane [Cantor, 1999]. The choice of protocell membrane is therefore important to provide an environment suitable for membrane proteins integrated into the protocell surface to function. It has been reported that membrane proteins (i.e. OmpF, LamB and FhuA [Stoenescu et al., 2004]) can be integrated within the membrane of polymersomes while maintaining their functionality.

The protocells made of polymersomes will be manufactured using a microfluidic platform. A miniature production and assembly line in the form of a self-contained microfluidic platform should be capable of producing protocells from purified components with the high efficiency and reproducibility that can be obtained with microfluidics. The application of microfluidics in synthetic biology for design, optimization and component production has been recently reviewed [Huang et al., 2014 ; Gulati et al., 2009].

It is well established that life possesses the unique ability to evolve in the face of selection pressures. Artificial protocells with their lack of replication machinery, do not posses this, as the ability to reproduce or divide is a condition for evolution to occur. In applied biotechnology the inability of protocells to evolve would be considered a merit, preventing undesired changes in the design. The downside is however that protocells cannot respond to applied selection pressures to optimize performance [Amezaga et al., 2014].

The optimal protocell must therefore be designed and pre-tested, it would be unrealistic to assume that simply integrating different parts will result in an optimally functional system. It will therefore be necessary to tag the proteins with fluorescent LOV domains, since genetically encoded fluorescent proteins (FPs) based on the flavin-binding LOV domain provide an advantage over green fluorescent protein (GFP). Recombinant expression of LOV domains in E. coli is fast and easy to detect given the innate ability of LOV domains to acquire their ubiquitous organic cofactor from the cellular environment [Christie, 2012]. Then it will be possible to use a microfluidic fluorescence sorter to separate protocells that have positioned the proteins in the optimal orientation.

A range of parameters such as concentrations, surfactants, flow rates etc. will then be systematically optimised through the application of a molecular evolution strategy recently pioneered. A synthetic organic reactor system using real-time in-line NMR spectroscopy can be utilized to self-optimize a reaction, exploiting feedback from the spectroscopic measurements, leading to a self-optimizing continuous-flow process. This real-time analytics enables the fast generation of data and information about the process studied and the fast optimization of the process, allowing direct control of the reactor, reagent inputs, and process conditions [Sans et al., 2015].

Sensory function of different synthetic transport modules such as NaChBac, from *Bacillus halodurans* [Ren et al., 2001] or $Na_V$PZ from *Paracoccus zeaxanthinifaciens* [Koishi et al., 2004] will be tested by measuring pH profiles of the artificial cells in response to treatment with a range of chloride and sodium concentrations. The affectivity and functionality of the artificial cells will be benchmarked against in vivo bacterial systems such as the previously designed recombinant *E.coli,* functionally expressing the same synthetic membrane transport modules [Amezaga et al., 2014].

In conclusion, if the above methodologies are successful, then a synthetic system in form of a self-sufficient light empowered protocell has been designed.

## DISCUSSION

Part of a good design remit is to choose a minimal system that will function as required. Protocells with well-defined synthetic applications, such as pharmaceutical synthesis, could be designed as stripped-down models, incorporating essential biosynthetic and bioenergetic pathways, but omitting a number of other complex processes required for "life" [Minas et al., (2015].

This review article discuss specific example of recombinant cell and protocell technology, moving from what is presently known to the suggestions of how novel application of existing methodologies could be utilized to design a complex synthetic system in form of a light empowered protocell for bioaccumulation and bio-diagnostics [Minas et al., (2015].

Protocells, with their polymersomes membranes serve a dual purpose, since firstly, they are in general more stable in the circulation than liposomes [Discher et al., 1999; Lee et al., 2011]. The possibility to load therapeutic molecules such as drugs, enzymes, peptides, and DNA and RNA fragments into polymersomes has been highlighted for a number of applications, making them very attractive vesicles for various applications in drug delivery, biomedical imaging, diagnostics, and biotechnology [Lomas et al., 2007; Kim et al., 2010; Massignani et al., 2010].

Secondly, as discussed so far, the inside-negative electrochemical membrane potential generated by $Cl^-$ pump activity via halorhodopsin could also be applied to drive the accumulation of cations, positively charged substances, into the protocell [Amezaga et al., 2014]. Thus, the expression of halorhodopsin could energize proton-coupled uptake of not only substances such as $Na^+$, but also hypothetically $Ca^{2+}$ (NaChBac selectivity can be converted from $Na^+$ to $Ca^{2+}$ by replacing an amino acid adjacent to glutamatic acid in the putative pore domain by a negatively charged aspartate [Yue et al., 2002]), and along the same tangent $Mg^{2+}$, $K^+$, $Fe^{2+}$, or toxic metals such as $Cd^{2+}$ (Tolypothric tenuis has demonstrated adsorption and removal of cadmium over wide pH and temperature ranges [Inthorn et al., 1996]), $Ni^{2+}$, or even a disease-related peptide through a selective cotransport protein such as for instance the highly expressing ion channel protein, NaChBac for bioaccumulation.

There is much ground to be explored before functional protocells become a reality. The discipline has still not yet progressed to the point where a fully functional self-replicating cell can be constructed. It is likely to be quite some time before a true autonomous synthetic cell can be designed, whereas the design and construction of simpler machines built for an array of important synthetic or bio-delivery purposes is almost within reach [Minas et al., (2015]. Further, a process to separate the protocells from the environment will also be necessary, otherwise the substances or metals will eventually be returned to where they came. The future goal will be to develop a generic solution which should lead to the design of a minimal protocell with the potential applications to report on inorganic and organic substances in water, soil or body fluids and thereby open a window of varied opportunities for applications in molecular medicine, agriculture and environmental water treatment.